\documentclass[aps,twocolumn]{revtex4}
\usepackage{graphicx}
\begin{document}
\newcommand{\D}{\displaystyle}
\title{Local density of states induced by anisotropic 
impurity scattering in a d-wave superconductor}  
\author{P. Pisarski and G. Hara\'n}
\affiliation{Institute of Physics, Politechnika
Wroc{\l}awska, Wybrze\.ze Wyspia\'nskiego 27, 50-370 Wroc{\l}aw,
Poland}

\begin{abstract}
We study a single impurity effect on the local density of states 
in a d-wave superconductor accounting for the momentum-dependent  
impurity potential. We show that the anisotropy of the scattering 
potential can alter significantly the spatial dependence of the 
quasiparticle density of states in the vicinity of the impurity.
\end{abstract}
\maketitle

\noindent
A search for the symmetry of the cuprate superconducting state involves 
among others theoretical studies of the disordered systems \cite{1,2,3}.  
The most direct probe of a simple defect impact on superconductivity 
is provided by the scanning tunneling microscopy (STM) measurement of the 
position-dependent quasiparticle density of states. 
The STM images of Zn and Ni substitutions at the  
planar Cu sites in ${\rm Bi_2Sr_2CaCu_2O_{8+\delta}}$ reveal a distinct 
four-fold symmetry of the local density of states (LDOS) \cite{4,5} 
predicted for the d-wave superconductor response to disorder \cite{6,7}.  
In addition to providing detailed information on the superconducting state,  
this kind of experiment may shed light on the nature of the quasiparticle 
scattering centers. Although the main feature of the four-fold symmetry 
of the tunneling currents is captured by the model of isotropic impurity 
scattering the observed spatial dependence of the quasiparticle density 
of states is far more complex, and may originate from a non trivial 
structure of the impurity potential. This aspect of the quasiparticle 
scattering process has been raised in the discussion of macroscopic 
measurements in these compounds \cite{8,9,10,11}.  
It has been shown that the unexpected for the d-wave 
superconductivity weak impurity-induced suppression of the critical 
temperature can be understood within a scenario of a momentum-dependent 
(anisotropic) impurity scattering potential \cite{12,13,14,15}.  
In the present paper we discuss the effect of anisotropic impurity potential 
on the local density of states in the vicinity of a single impurity 
and verify to what extend this scenario can reproduce the 
STM maps of ${\rm Bi_2Sr_2CaCu_2O_{8+\delta}}$. There are studies of 
the momentum-dependent impurity scattering which indicate the existence of 
the impurity-bound states even for the potential strength far from 
resonant \cite{16}. Therefore, we focus here on the symmetry of the LDOS 
and its spatial dependence. For this purpose and for analytical simplicity 
we consider scattering in the Born approximation. Hereafter we show that 
the anisotropy of the scattering potential alters the local density of 
states in the d-wave superconductor but preserves its tetragonal symmetry. 
Compared to the effect of the isotropic impurity potential it enhances the 
spatial variation of the quasiparticle density of states and can lead to a 
long-range spatial modulation of the LDOS spectra. We note that the anisotropy 
of the impurity potential, if the same as the anisotropy of the superconducting 
order parameter, enhances the relative magnitude of oscillations in the LDOS for 
the direction of a gap maximum, especially in the vicinity of the impurity, 
whereas the impurity potential with maxima in the nodal region inverts this 
tendency and leads to a larger quasiparticle density of states along the gap 
node direction. 
 
We consider the momentum-dependent impurity potential   
\cite{12,13,14,15,15a} 

\begin{equation}
\label{e1}
\hat{v}\left({\bf k},{\bf k}'\right)=\left[v_i+
v_af\left({\bf k}\right)f\left({\bf k}'\right)\right]\hat{\tau}_3
\end{equation}

\noindent
where $v_i$, $v_a$ are isotropic and anisotropic scattering amplitudes, 
respectively, $f\left({\bf k}\right)$ is the anisotropy function, and 
$\hat{\tau}_i$ ($i=1,2,3$) are Pauli matrices.  
We assume $f\left({\bf k}\right)=\pm 1$ such as its Fermi surface (FS) 
average vanishes, i.e.,   
$\D\left<f\right>=\int_{FS}dS_k n({\bf k})f\left({\bf k}\right)=0$, where 
$n({\bf k})$ is the normalized angle resolved FS density of states, 
$\D\int_{FS}dS_k n({\bf k})=1$. Potential determined by 
$f\left({\bf k}\right)=sgn\left(k^2_x-k_y^2\right)=sgn\left(cos2\phi\right)$,  
which is in phase with the d-wave superconducting order parameter 
leads to a particularly moderate suppression of the critical temperature 
\cite{12,15,15a}. In the above $\phi$ is a polar angle.  

We study the effect of the impurity potential (\ref{e1}) on the quasiparticle 
states using the one particle Green's function of the $d_{x^2-y^2}$-wave 
superconducting state in the particle-hole notation 

\begin{equation}
\label{e2}
\hat{G}_0\left({\bf k},\omega_n\right)=\left[i\omega_n\hat{\tau}_0
-\xi_{\bf k}\hat{\tau}_3-\Delta\left({\bf k}\right)\hat{\tau}_1\right]^{-1}
\end{equation} 

\noindent
where $\Delta({\bf k})=\Delta\left(k^2_x-k^2_y\right)=\Delta cos2\phi$ 
represents the superconducting order parameter, $\omega_n$ is the 
Matsubara frequency, $\xi_{\bf k}=\varepsilon_{\bf k}-\varepsilon_F$ 
is the quasiparticle energy in the normal state, $\varepsilon_F$ 
is the Fermi energy, and 
$\hat{\tau}_0$ is the identity matrix. For simplicity we assume a parabolic 
dispersion and a two-dimensional (planar) superconductivity. Neglecting 
the anisotropy of the Fermi surface we concentrate on the feature of coupled 
anisotropies of the order parameter and the impurity potential. 
The impurity effect is studied at zero temperature by the analytic 
continuation of (\ref{e2})  

\begin{equation}
\label{e4}
\hat{G}_0\left({\bf k},\omega\right)=
\hat{G}_0\left({\bf k},\omega_n\right)_{\D i\omega_n=\omega+i0^{+}}
\end{equation}

\noindent
In order to discuss the real space distribution of the quasiparticle 
states we take a two-dimensional Fourier transform of the Green's 
function (\ref{e4}), which for the parabolic band 
\mbox{$\varepsilon_{\bf k}=k^2/2m$} and for positive $\omega$ smaller than 
the Fermi energy reads 

\begin{widetext}
\begin{equation}
\label{e5}
\D\hat{G}_0\left({\bf r},\omega\right)=\int\limits^{\D\infty}_{\D -\infty}
\frac{\D d^2 k}{\D\left(2\pi\right)^2}e^{\D i{\bf kr}}
\hat{G}_0\left({\bf k},\omega\right)=
\frac{\D -i\pi m}{\D 2\left(2\pi\right)^2}
\int\limits^{\D 2\pi}_{\D 0}d\phi_{\bf k} 
\left[\frac{\D\omega\hat{\tau}_0+\Delta\left({\bf k}\right)\hat{\tau}_1}
{\D\sqrt{\omega^2-\Delta^2\left({\bf k}\right)}}
\left(e^{\D i{\bf k_{+}r}}+e^{\D -i{\bf k_{-}r}}\right)
+\hat{\tau}_3\left(e^{\D i{\bf k_{+}r}}-e^{\D -i{\bf k_{-}r}}\right)\right]
\end{equation}
\end{widetext}

\noindent
where similarly to the three-dimensional case \cite{17} 
\mbox{$k_{\pm}=\sqrt{2m}\left[\varepsilon_F\pm
\sqrt{\omega^2-\Delta^2\left({\bf k}\right)}\right]^{1/2}$} and 
$Im\sqrt{\omega^2-\Delta^2\left({\bf k}\right)}\ge 0$.  
For negative $\omega$ one must multiply the imaginary part of (\ref{e5})  
by $sgn\left(\omega\right)$. Here $\hbar=1$.  

We consider the impurity effect in the Born approximation, that is, proceed 
with the perturbative expansion of the Green's function up to the second 
order term in the scattering potential $\hat{v}$. The linear correction 
results from the impurity-induced change of the Fermi energy 

\begin{equation}
\label{e6}
\D\delta \hat{G}'\left({\bf k},{\bf k}',\omega\right)=
\hat{G}_0\left({\bf k},\omega\right)
\hat{v}\left({\bf k},{\bf k}'\right)
\hat{G}_0\left({\bf k}',\omega\right)
\end{equation}

\noindent
while the quadratic term represents the proper Born correction 
and takes the finite quasiparticle life-time into account 

\begin{equation}
\label{e7}
\D\delta \hat{G}''\left({\bf k},{\bf k}',\omega\right)=
\hat{G}_0\left({\bf k},\omega\right)
\hat{\Sigma}_B\left({\bf k},{\bf k}'\right)
\hat{G}_0\left({\bf k}',\omega\right)
\end{equation}

\noindent
where the self-energy $\hat{\Sigma}_B$ is determined by  

\begin{equation}
\label{e8}
\hat{\Sigma}_B\left({\bf k},{\bf k}'\right)=\sum_{\bf k''}
\hat{v}\left({\bf k},{\bf k''}\right)\hat{G}_0\left({\bf k''},\omega\right)
\hat{v}\left({\bf k''},{\bf k}'\right)
\end{equation}

\noindent
Both corrections are evaluated for a single impurity which means that the 
self-energy is not determined self-consistently. We have neglected changes 
to the absolute value of the order parameter in the vicinity of the 
impurity which will not change the symmetry of LDOS images but their 
absolute magnitudes. The total correction to the Green's function 
reads 

\begin{equation}
\label{e9}
\delta\hat{G}\left({\bf k},{\bf k}',\omega\right)=
\delta\hat{G}'\left({\bf k},{\bf k}',\omega\right)+
\delta\hat{G}''\left({\bf k},{\bf k}',\omega\right)
\end{equation}

\noindent
and its real space transform is given by 

\begin{equation}
\label{e10}
\D\delta \hat{G}\left({\bf r},\omega\right)=\sum_{{\bf k},{\bf k}'}
e^{\D i\left({\bf k}-
{\bf k}'\right){\bf r}}\delta \hat{G}\left({\bf k},{\bf k}',\omega\right)
\end{equation}

\noindent
Function (\ref{e10}) determines the position dependent impurity-induced change 
of the quasiparticle density of states near a single impurity 

\begin{equation}
\label{e11}
\D\delta N\left({\bf r},\omega\right)=
-\frac{\D 1}{\D\pi}Im\left\{\delta G_{11}\left({\bf r},\omega\right)\right\}
\end{equation}

\noindent
We have evaluated the LDOS for equal impurity scattering strength in isotropic 
and anisotropic channel $\pi N_0v_i=\pi N_0v_a=0.1$, where $N_0$ is the 
density of states per spin at the Fermi level in the normal state, 
and for the coherence 
length $\xi_0=12\pi k_F^{-1}$. The distance scale in the figures is set by 
$k_F^{-1}$. We observe a pronounced spatial variation of the quasiparticle density 
of states for anisotropic impurity potential. This feature is in agreement 
with the result of an enhanced LDOS due to a finite range of the scattering 
potential \cite{18}. In Figs. 1a-c we show the distance dependence of the 
impurity-induced change to the LDOS at the quasiparticle energy 
$\omega=1.1\Delta$ along the a-axis ($\phi=0$), i.e., gap maximum, and nodal 
($\phi=\pi/4$) direction for isotropic scattering (Fig. 1a) and two representative 
anisotropic impurity potentials (\ref{e1}): 
$f\left({\bf k}\right)=sgn\left(k^2_x-k_y^2\right)=sgn\left(cos2\phi\right)$ 
which is in phase with the d-wave order parameter (Fig. 1b), and 
$f\left({\bf k}\right)=sgn\left(k_xk_y\right)=sgn\left(sin2\phi\right)$ 
(Fig. 1c) scattering out of the order parameter phase. We note that the 
impurity potential in phase with the order parameter enhances the magnitude 
of oscillations in the LDOS for the direction of a gap maximum, especially 
in the vicinity of the impurity, whereas the out of phase impurity potential 
inverts this tendency and leads to a larger quasiparticle density of states 
in the nodal regions. Significant is also very weak anisotropy of the LDOS 
near the isotropic impurity (Fig. 1a) where discernible differences between 
$\phi=0$ and $\phi=\pi/4$ directions occur at a distance of about 
$50k_F^{-1}$ from impurity \cite{6}. It shows that visible spatial 
distribution of intensity maxima and minima in the STM maps is induced by the 
anisotropy of defect potential. Therefore, we suggest that the apparent anisotropy  
of the experimental images \cite{4,5} may possibly result in part from the presence 
of the momentum-dependent scattering potential. We have also performed the LDOS  
calculation for quasiparticles below the gap threshold energy and displayed 
them for $\omega=0.1\Delta$ in Figs. 2a-c. Another interesting feature of the 
anisotropic scattering is a presence of a long-range spatial modulation of the 
LDOS for the impurity potential in phase with the d-wave order parameter, i.e., 
$f\left({\bf k}\right)=sgn\left(k^2_x-k_y^2\right)$. Such a modulation is 
absent for isotropic or out of phase scattering (Fig. 3). We summarize our 
discussion by showing in Figs. 4-5 comprehensive pictures of the STM images at 
the frequencies $\omega=1.1\Delta$ and $\omega=0.1\Delta$ around simple defects 
of isotropic and anisotropic Born potential in the ab plane of the d-wave 
superconductor. The potential strength, coherence length and distance 
units have been fixed as in Figs. 1-3. The quasiparticle density of states is 
given in the units of the FS two-spin density of states $N(0)$, $N(0)=2N_0$, 
and varies from the lowest (black) to the highest (white) value according to 
the scale in each figure. We have checked that similarly to the effect 
on the critical temperature \cite{13} any other scattering potential 
which is orthogonal to the order parameter(in the sense of the FS integral as 
the scalar product) gives qualitatively the same LDOS as the discussed above 
$f\left({\bf k}\right)=sgn\left(sin2\phi\right)$, i.e., out of phase potential.  
The effect of potentials in phase with the order parameter like 
$f\left({\bf k}\right)=cos2\phi$ agrees qualitatively with 
$f\left({\bf k}\right)=sgn\left(cos2\phi\right)$. Therefore, both studied 
anisotropies can be considered representative for the impurity potential 
(\ref{e1}). Although we have restricted our study to nonmagnetic impurity 
scattering, as a closing remark we note that inclusion of spin $S=1/2$, $1$ or 
$3/2$ scattering does not contribute any quantitative change to the results 
for the impurity potential obeying the Born approximation.\\ 
Concluding, we have shown that the anisotropy of the impurity potential enhances 
the quasiparticle LDOS in the vicinity of the impurity and leads to discernible  
differences of its intensity in horizontal and diagonal direction. Depending on 
its symmetry the impurity potential can rotate the LDOS maxima and minima by 
$45^{\circ}$ degrees and can induce a long-range spatial modulation of the 
quasiparticle density of states but cannot change the induced by the d-wave 
superconductivity four-fold symmetry of the STM images.\\ 
\indent
Note added in proof: A similar modulation to the one in Fig. 3(b) is seen 
along the gap nodal direction for the potential out of the order parameter 
phase.\\
\indent
We would like to acknowledge helpful discussions with Dr. L. Borkowski. 
The work was supported in part by KBN grant No. 5P03B05820.

\begin{figure}[p]
\parbox{1cm}{\vfill $$\frac{\delta N\left(r\right)}{N\left(0\right)}
\vspace{5ex}$$\vfill}
\parbox{5cm}{\includegraphics[height=5cm,width=5cm]{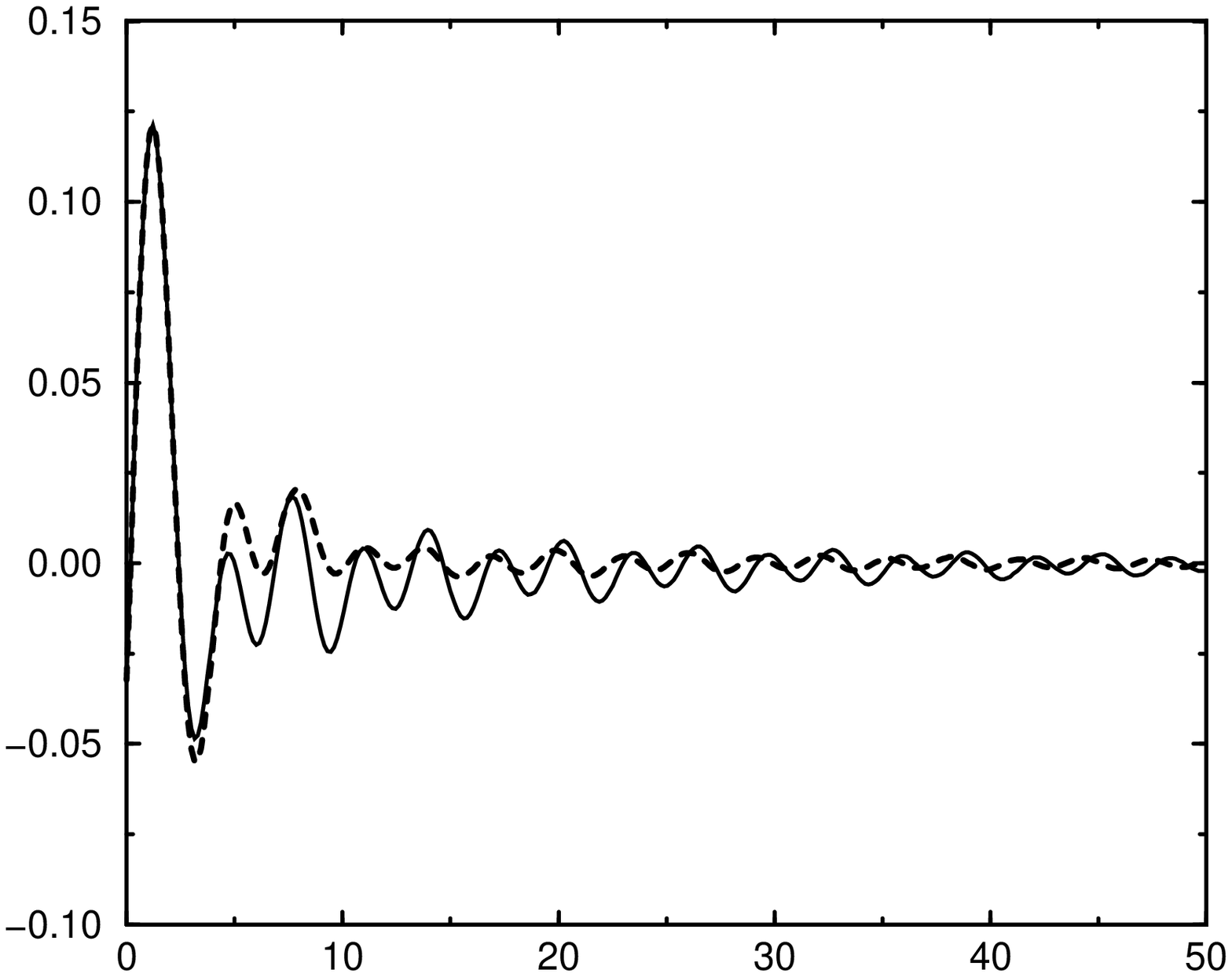}}
\parbox{0.5cm}{\hfill}
\parbox{9.5cm}{\vspace{-6ex}\hfill $$r\left[k^{-1}_F\right]$$\hspace*{5em} (a)\hfill}
\end{figure}

\begin{figure}[p]
\parbox{1cm}{\vfill $$\frac{\delta N\left(r\right)}{N\left(0\right)}
\vspace{5ex}$$\vfill}
\parbox{5cm}{\includegraphics[height=5cm,width=5cm]{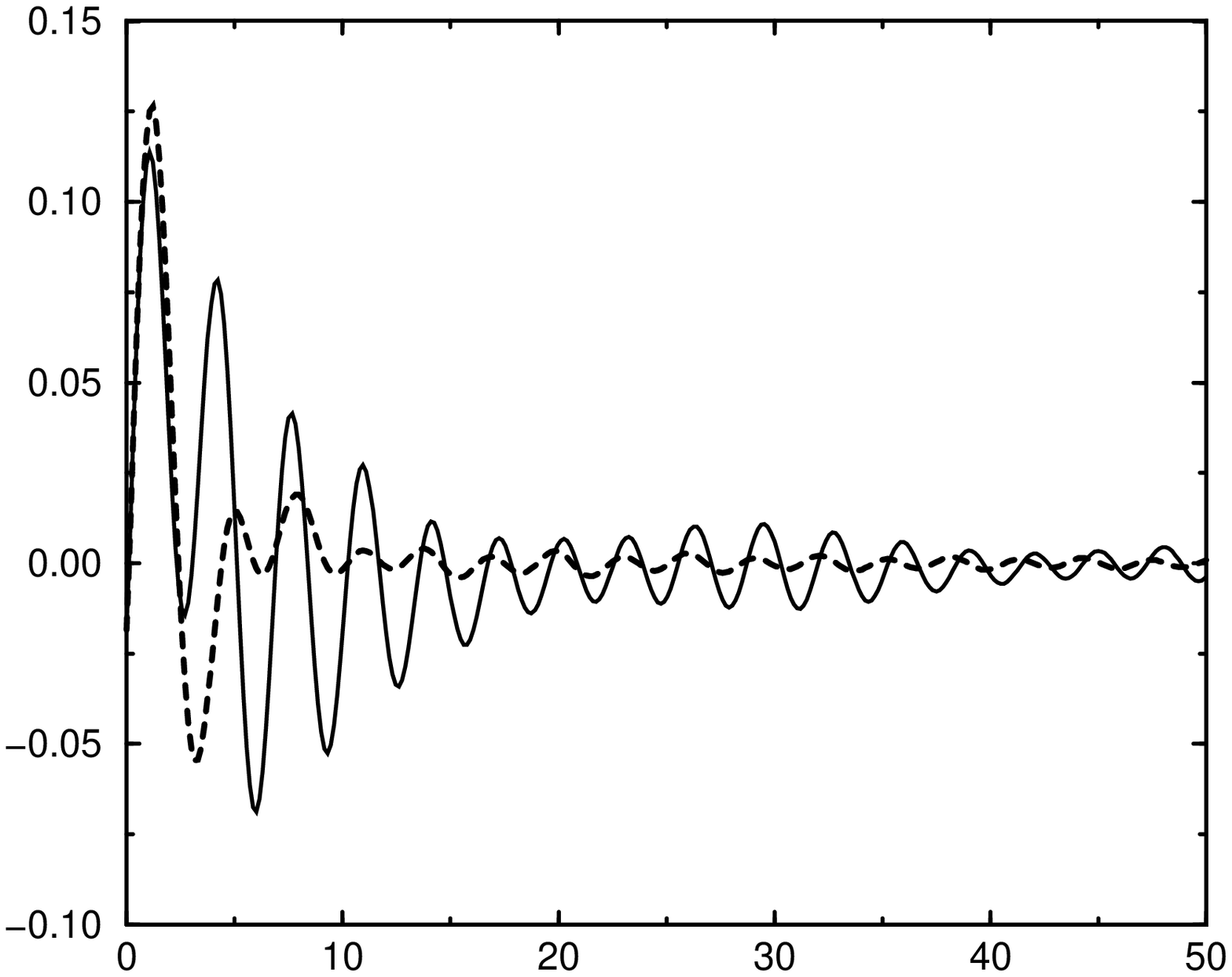}}
\parbox{0.5cm}{\hfill}
\parbox{9.5cm}{\vspace{-6ex}\hfill $$r\left[k^{-1}_F\right]$$\hspace*{5em} (b)\hfill}
\end{figure}

\begin{figure}[p]
\parbox{1cm}{\vfill $$\frac{\delta N\left(r\right)}{N\left(0\right)}
\vspace{5ex}$$\vfill}
\parbox{5cm}{\includegraphics[height=5cm,width=5cm]{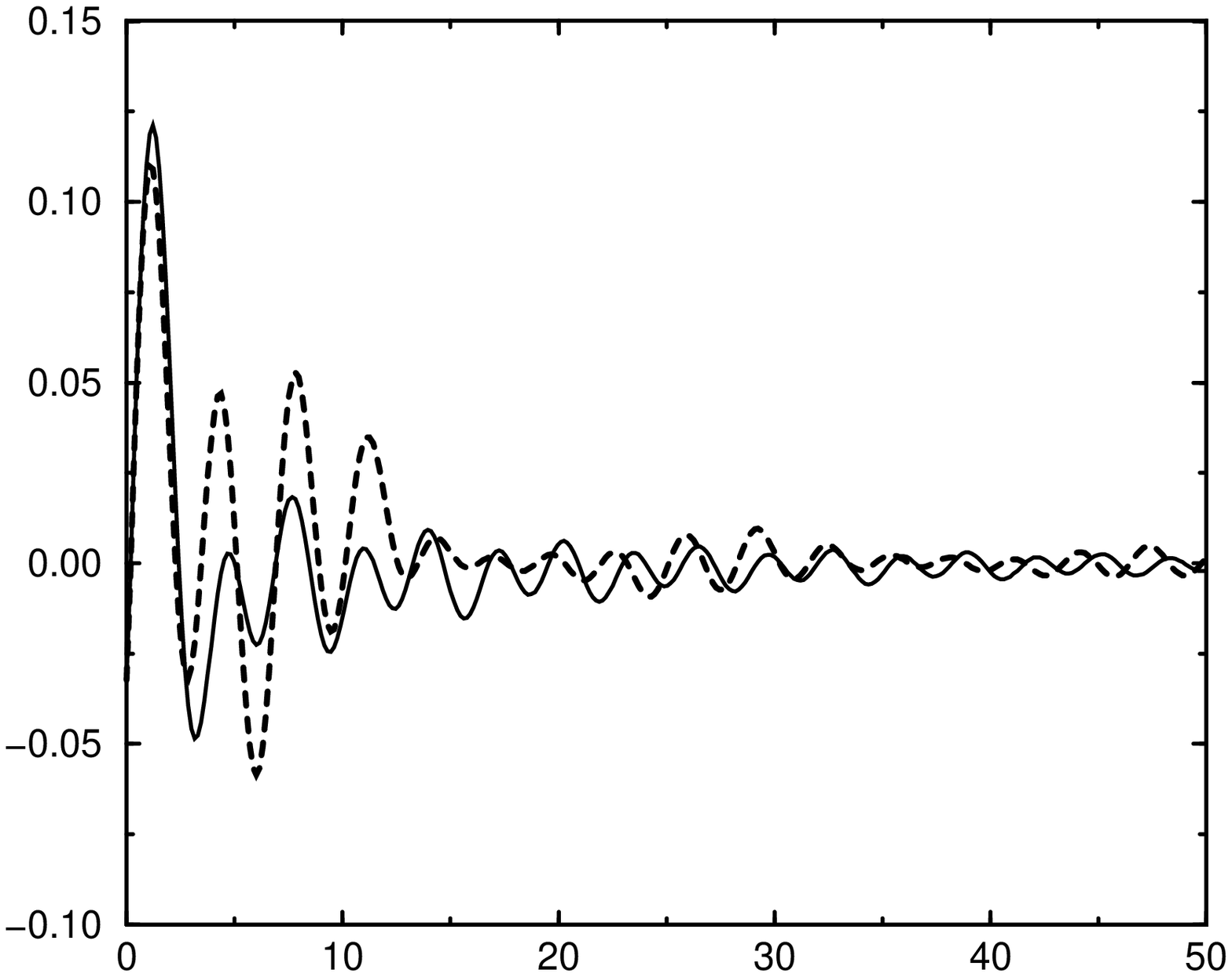}}
\parbox{0.5cm}{\hfill}
\parbox{9.5cm}{\vspace{-6ex}\hfill $$r\left[k^{-1}_F\right]$$\hspace*{5em} (c)\hfill}
\vspace*{0.5cm}
\caption{Distance dependence of the impurity-induced change to the LDOS,
$\delta N(r)$, at the quasiparticle energy $\omega=1.1\Delta$
along the a-axis ($\phi=0$ - solid line) and diagonal direction
($\phi=\pi/4$ - dashed line) away from the impurity located
at $r=0$ for
a) isotropic impurity potential;
b) potential in phase with the order parameter, $f({\bf k})=sgn(k^2_x-k^2_y)$;
c) potential out of the order parameter phase, $f({\bf k})=sgn(k_xk_y)$.
$N(0)$ is the FS density of states in the normal state and $k_F$ is the Fermi
momentum.}
\end{figure} 

\begin{figure}[p]
\parbox{1cm}{\vfill $$\frac{\delta N\left(r\right)}{N\left(0\right)}
\vspace{5ex}$$\vfill}
\parbox{5cm}{\includegraphics[height=5cm,width=5cm]{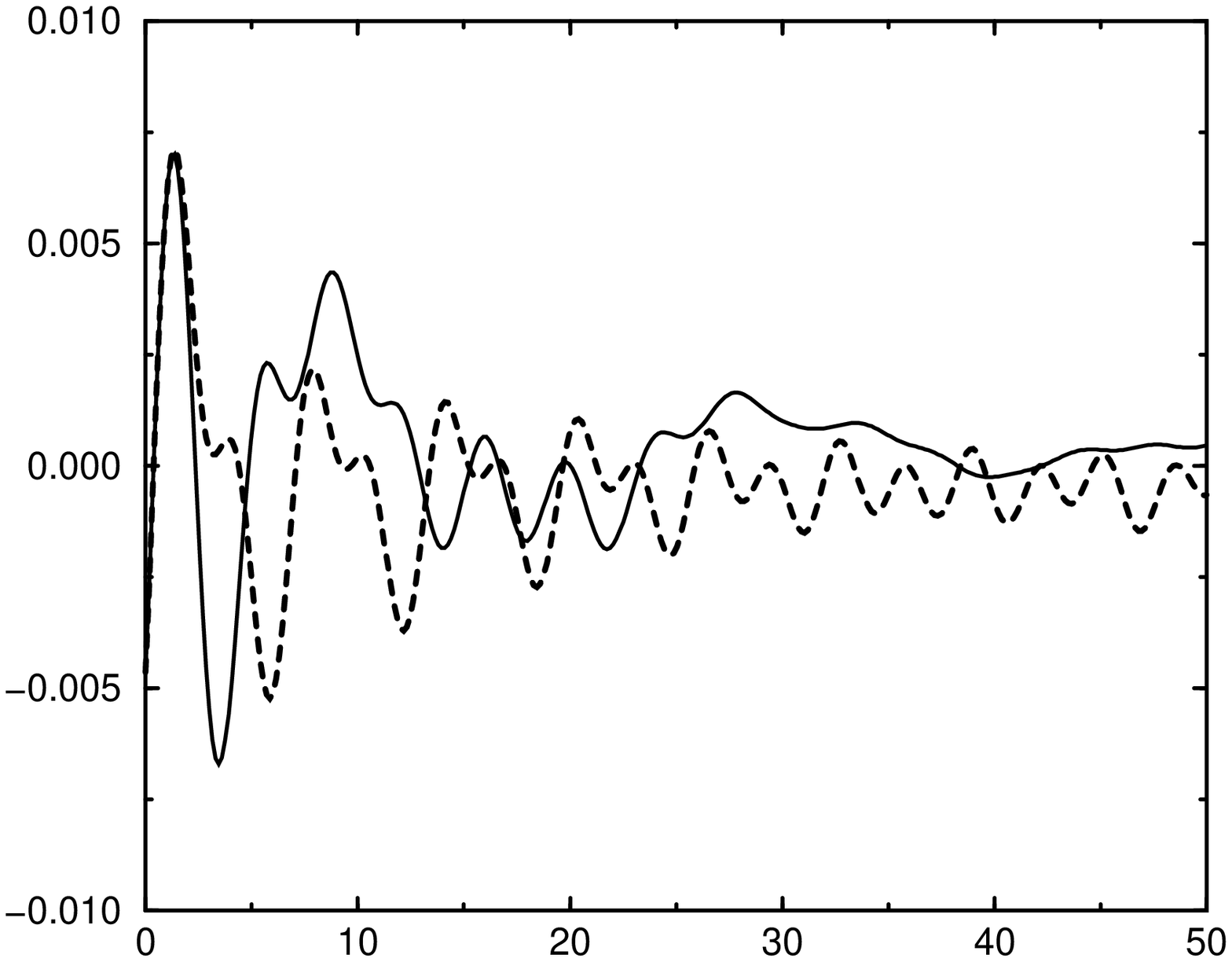}}
\parbox{0.5cm}{\hfill}
\parbox{9.5cm}{\vspace{-6ex}\hfill $$r\left[k^{-1}_F\right]$$\hspace*{5em} (a)\hfill}
\end{figure}

\begin{figure}[p]
\parbox{1cm}{\vfill $$\frac{\delta N\left(r\right)}{N\left(0\right)}
\vspace{5ex}$$\vfill}
\parbox{5cm}{\includegraphics[height=5cm,width=5cm]{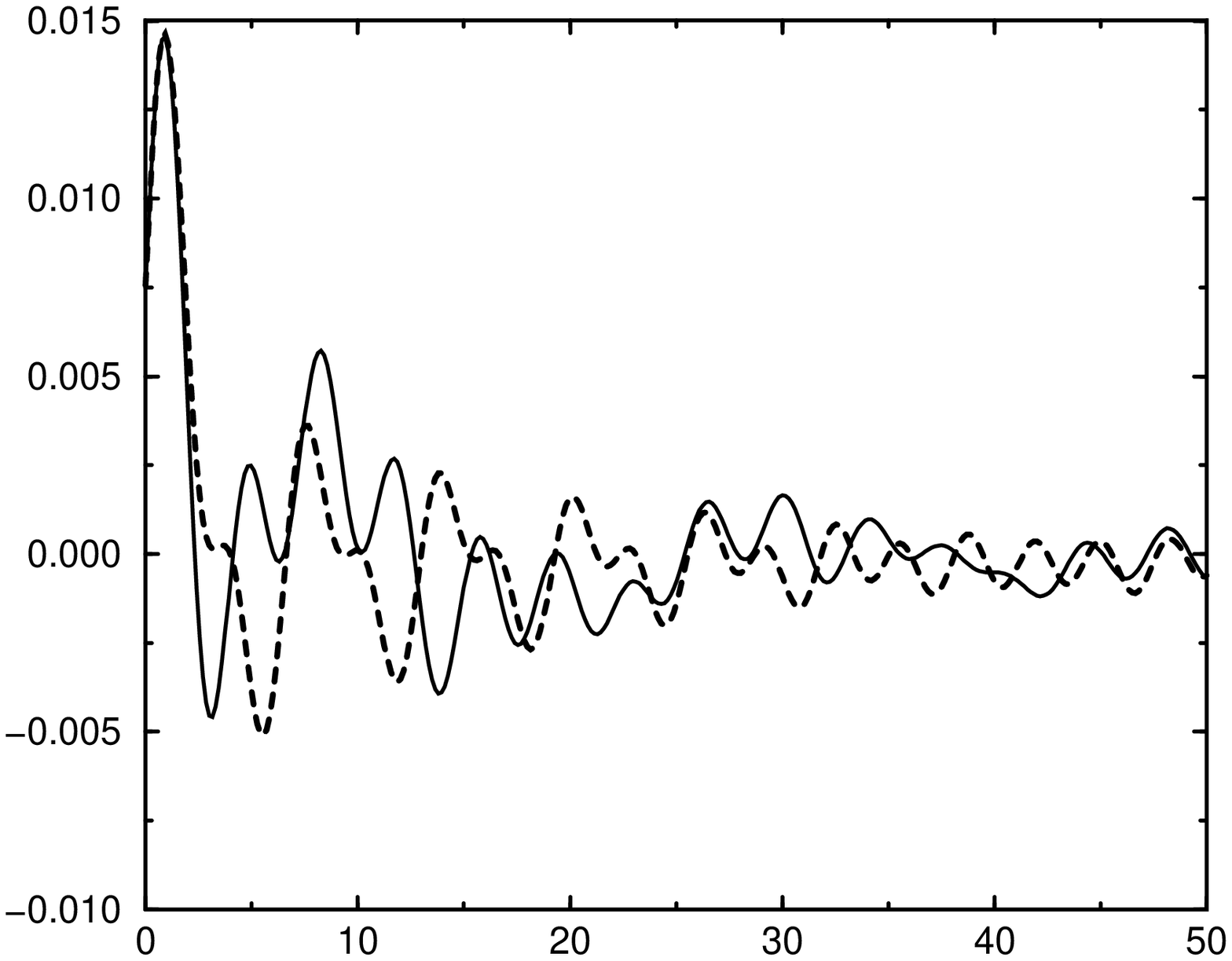}}
\parbox{0.5cm}{\hfill}
\parbox{9.5cm}{\vspace{-6ex}\hfill $$r\left[k^{-1}_F\right]$$\hspace*{5em} (b)\hfill}
\end{figure}

\begin{figure}[p]
\parbox{1cm}{\vfill $$\frac{\delta N\left(r\right)}{N\left(0\right)}
\vspace{5ex}$$\vfill}
\parbox{5cm}{\includegraphics[height=5cm,width=5cm]{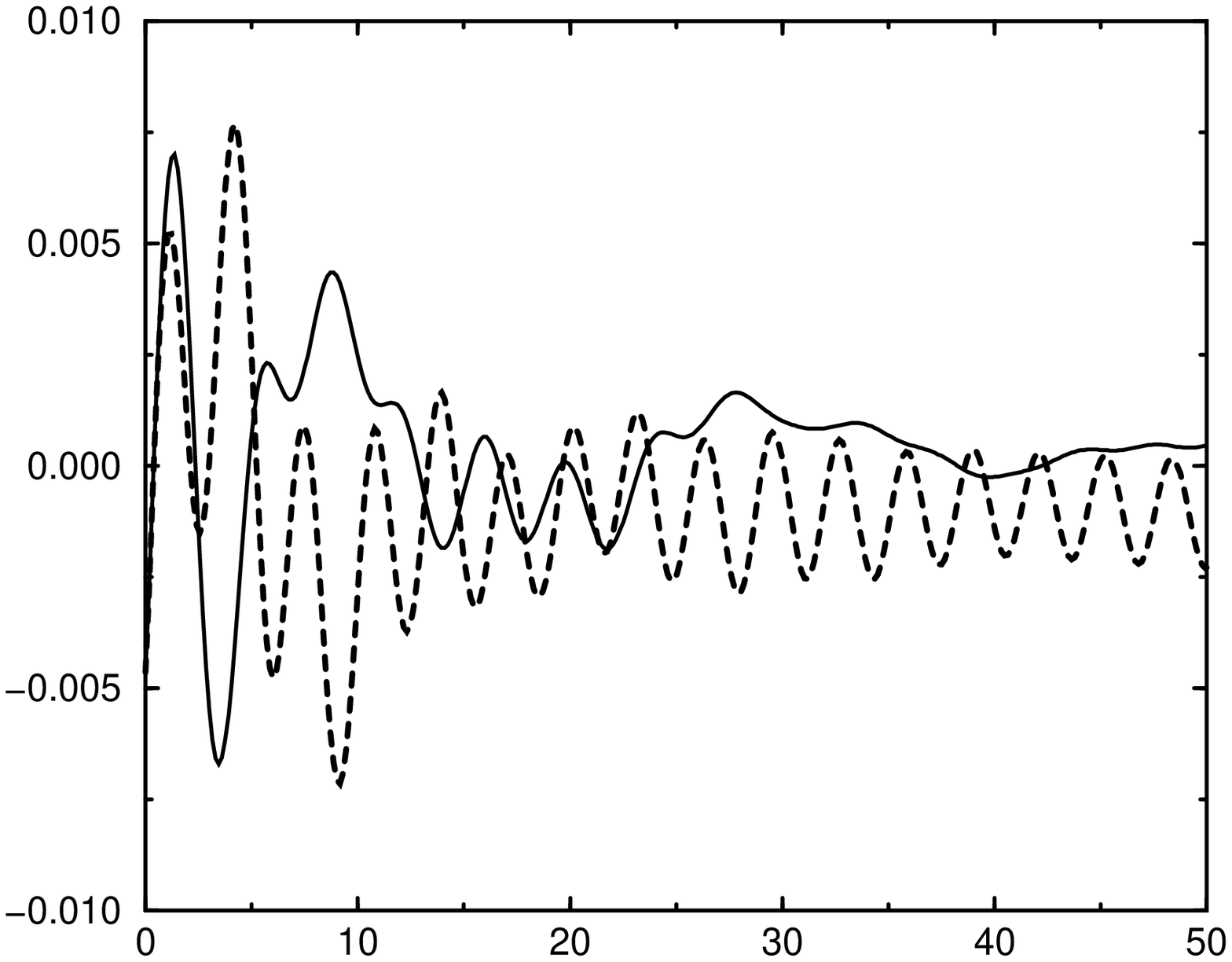}}
\parbox{0.5cm}{\hfill}
\parbox{9.5cm}{\vspace{-6ex}\hfill $$r\left[k^{-1}_F\right]$$\hspace*{5em} (c)\hfill}
\vspace*{0.5cm}
\caption{Distance dependence of the impurity-induced change to the LDOS
at the quasiparticle energy $\omega=0.1\Delta$
along the a-axis ($\phi=0$ - solid line) and diagonal direction
($\phi=\pi/4$ - dashed line) away from the impurity located
at $r=0$ for
a) isotropic impurity potential;
b) potential in phase with the order parameter, $f({\bf k})=sgn(k^2_x-k^2_y)$;
c) potential out of the order parameter phase, $f({\bf k})=sgn(k_xk_y)$.}
\end{figure}

\begin{figure}[p]
\parbox{1cm}{\vfill $$\frac{\delta N\left(r\right)}{N\left(0\right)}
\vspace{5ex}$$\vfill}
\parbox{5cm}{\includegraphics[height=5cm,width=5cm]{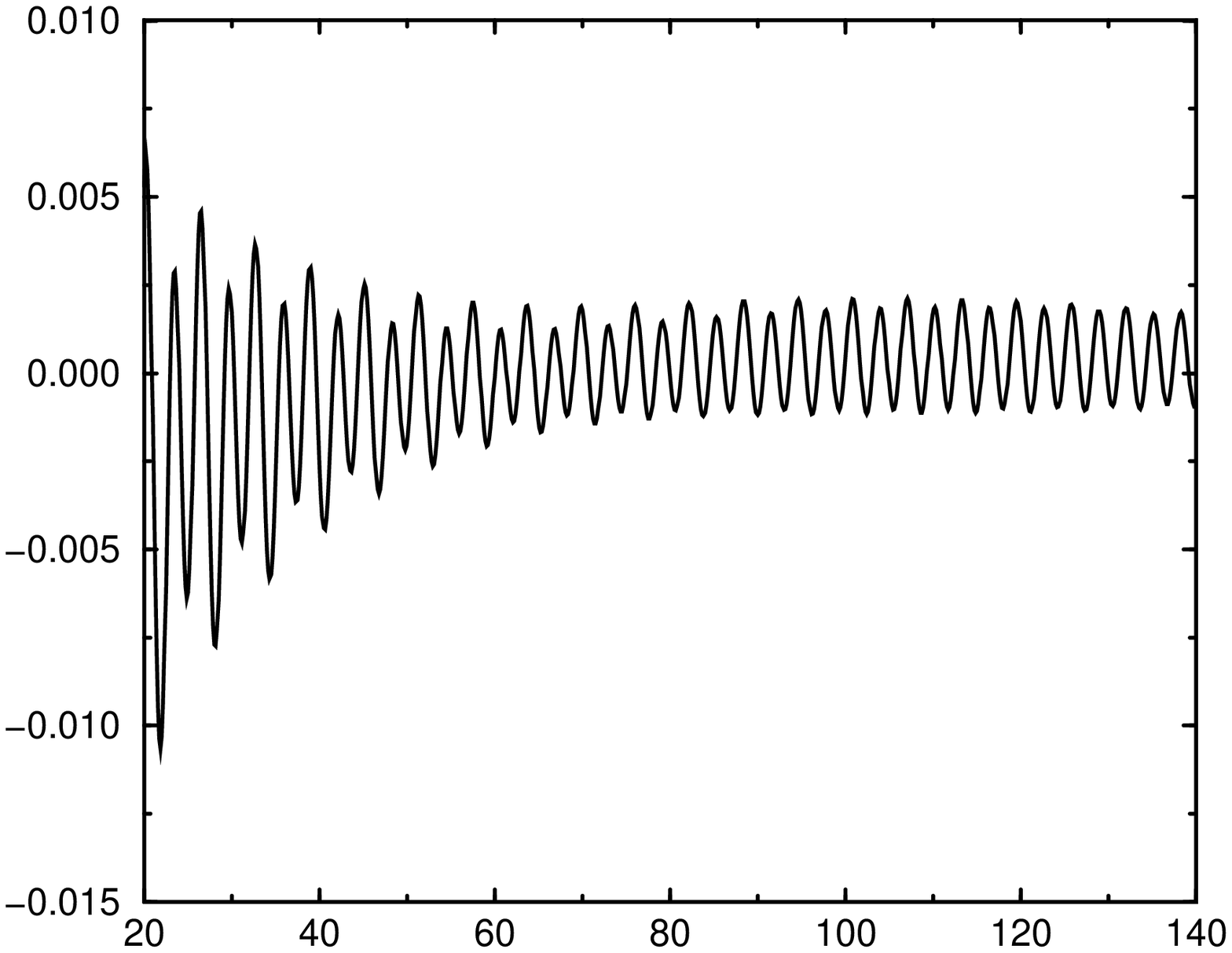}}
\parbox{0.5cm}{\hfill}
\parbox{9.5cm}{\vspace{-6ex}\hfill $$r\left[k^{-1}_F\right]$$\hspace*{5em} (a)\hfill}
\end{figure}

\begin{figure}[p]
\parbox{1cm}{\vfill $$\frac{\delta N\left(r\right)}{N\left(0\right)}
\vspace{5ex}$$\vfill}
\parbox{5cm}{\includegraphics[height=5cm,width=5cm]{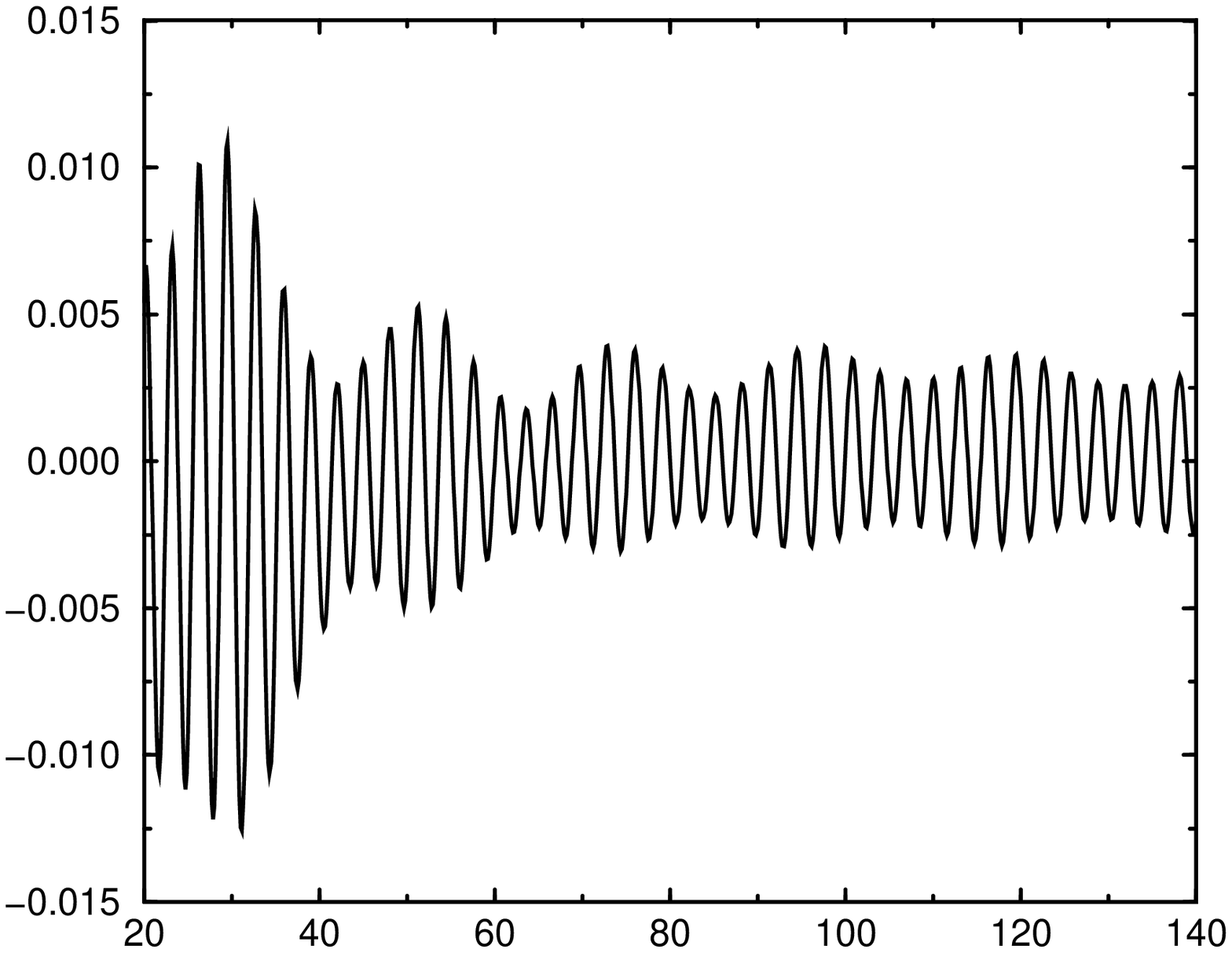}}
\parbox{0.5cm}{\hfill}
\parbox{9.5cm}{\vspace{-6ex}\hfill $$r\left[k^{-1}_F\right]$$\hspace*{5em} (b)\hfill}
\vspace*{0.5cm}
\caption{Long-range distance dependence of the impurity-induced change to the
quasiparticle density of states along $\phi=0$ direction
at the frequency $\omega=1.1\Delta$ for
a) isotropic impurity potential and the out of the order parameter phase
potential, $f({\bf k})=sgn(k_xk_y)$;
b) potential in phase with the order parameter, $f({\bf k})=sgn(k^2_x-k^2_y)$.\\
\vspace*{5cm}}
\end{figure}

\begin{figure}[p]
\includegraphics[height=4.92cm,width=6cm]{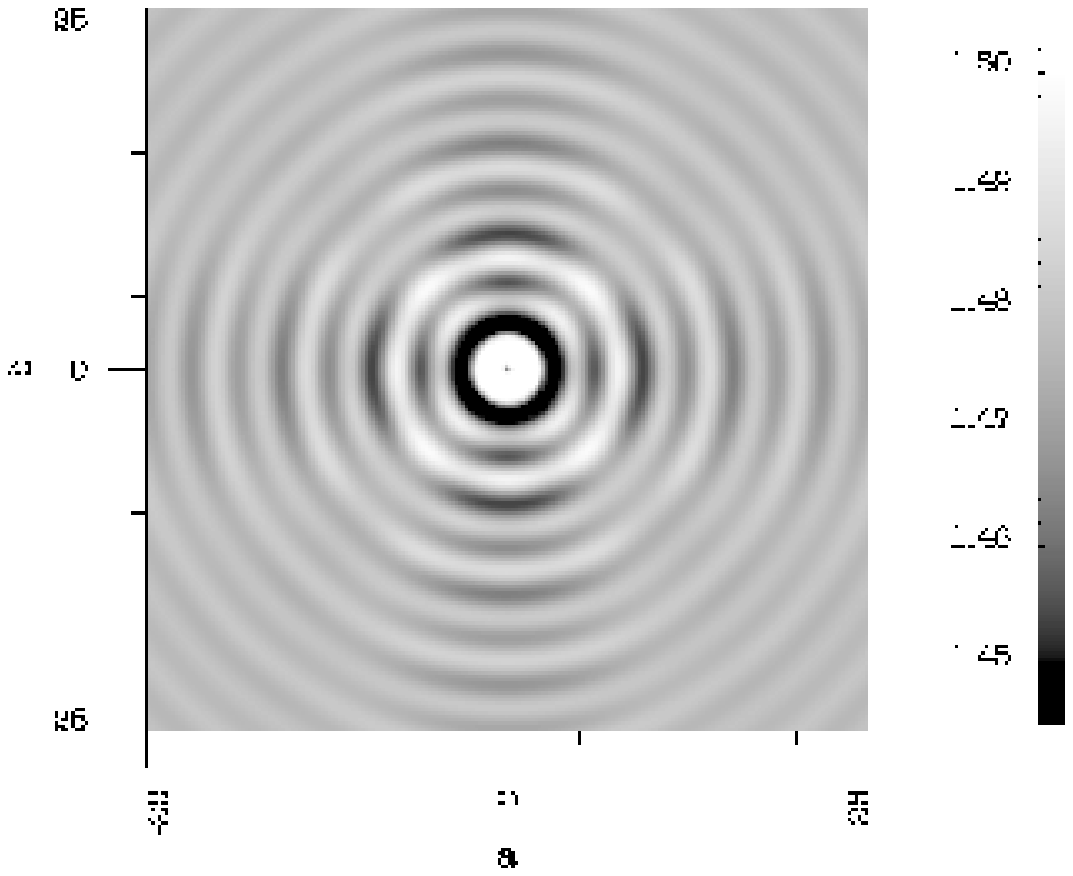}
\parbox{1cm}{\hfill\vspace{1cm} (a)\hfill}
\end{figure}

\begin{figure}[p]
\includegraphics[height=4.92cm,width=6cm]{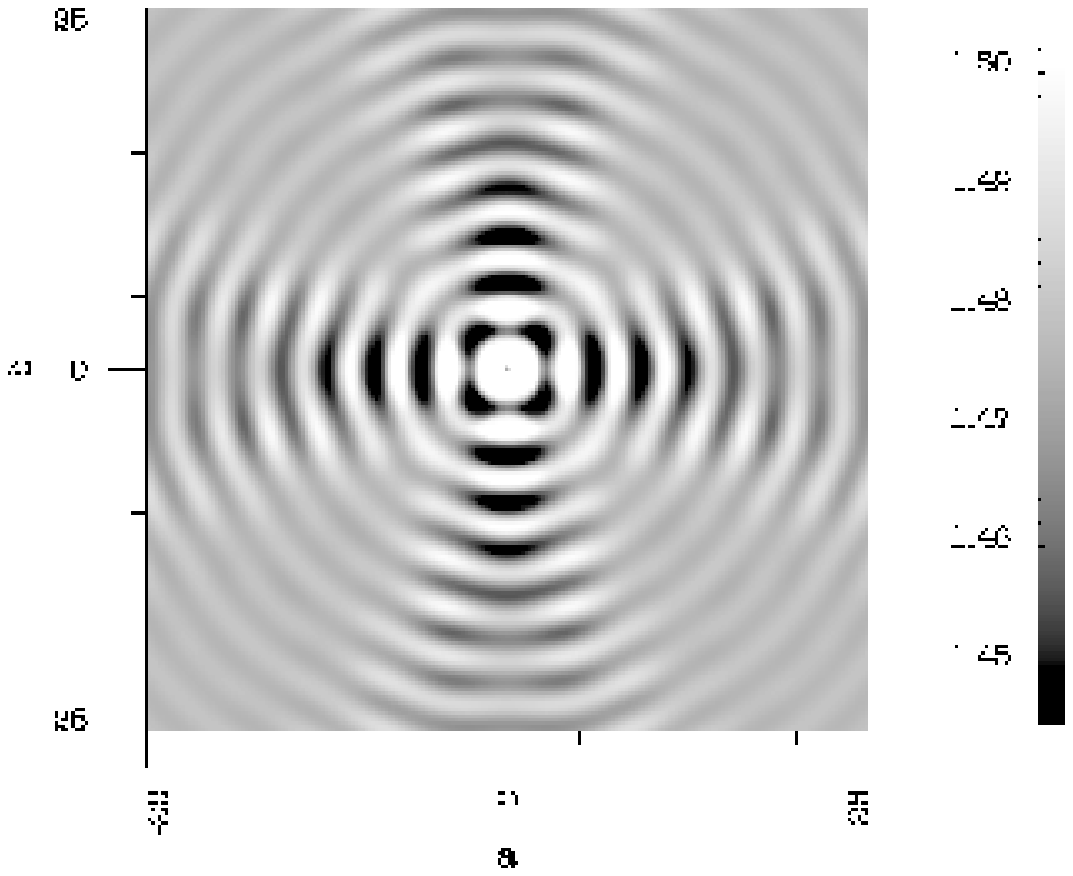}
\parbox{1cm}{\hfill\vspace{1cm} (b)\hfill}
\end{figure}

\begin{figure}[p]
\includegraphics[height=4.92cm,width=6cm]{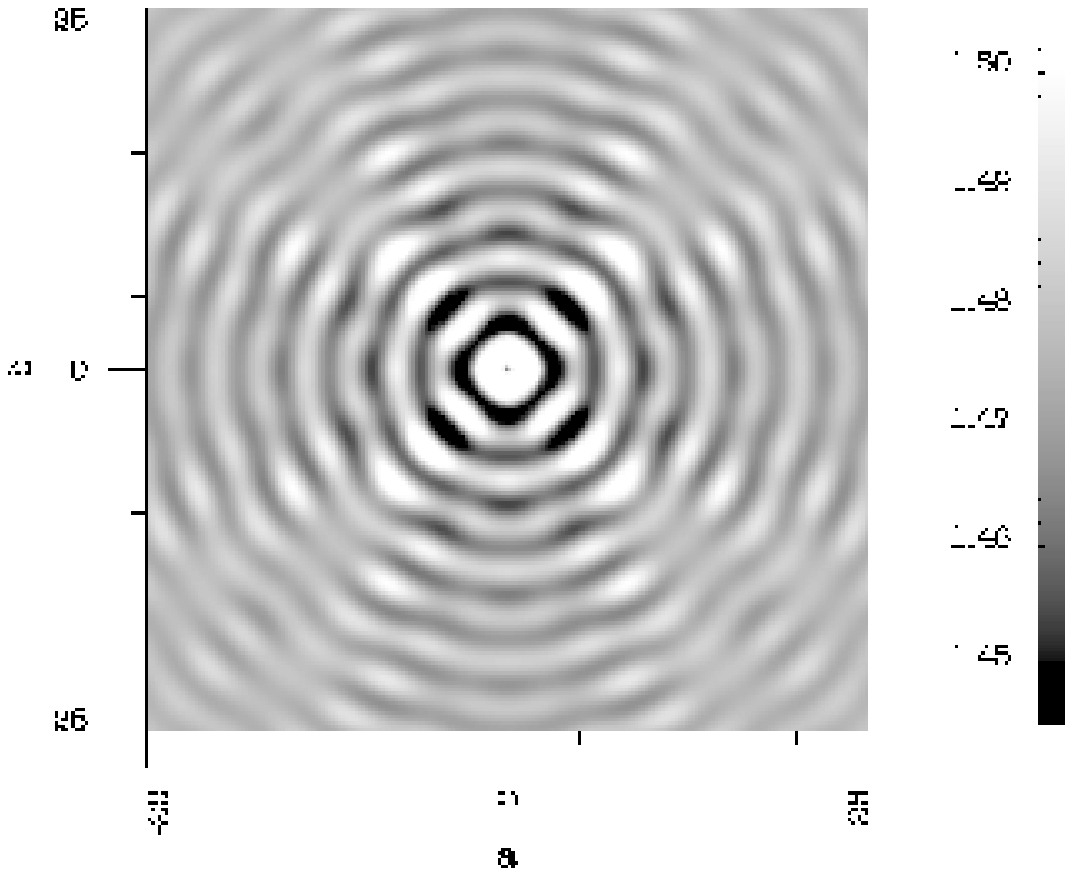}
\parbox{1cm}{\hfill\vspace{1cm} (c)\hfill}
\vspace*{0.5cm}
\caption{LDOS-map at the frequency $\omega=1.1\Delta$ around the Born impurity
located at (0,0) in the ab plane of the d-wave superconductor for
a) isotropic impurity potential;
b) potential in phase with the order parameter, $f({\bf k})=sgn(k^2_x-k^2_y)$;
c) potential out of the order parameter phase, $f({\bf k})=sgn(k_xk_y)$.
The density of states is given in the units of $N(0)$ by the scale next to each
map and the distance is measured in $k_F^{-1}$ units.}
\end{figure}

\begin{figure}[p]
\includegraphics[height=4.92cm,width=6cm]{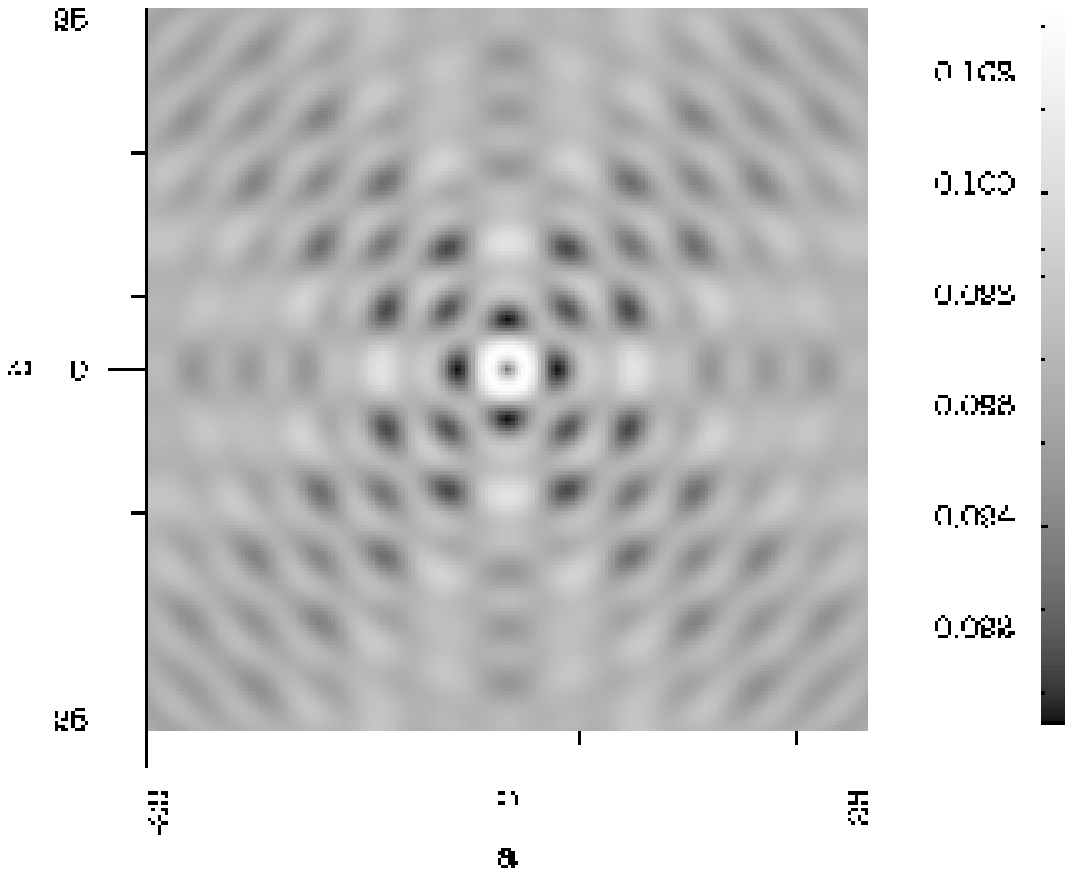}
\parbox{1cm}{\hfill\vspace{1cm} (a)\hfill}
\end{figure}

\begin{figure}[p]
\includegraphics[height=4.92cm,width=6cm]{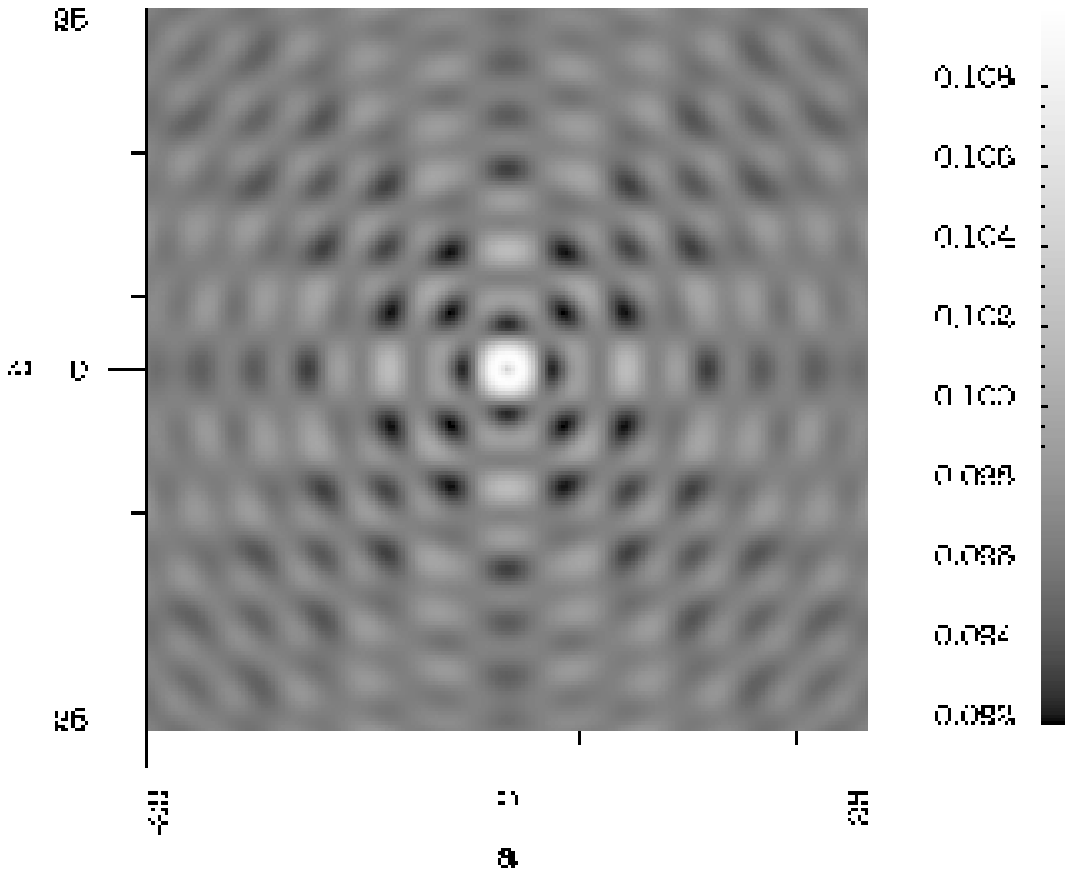}
\parbox{1cm}{\hfill\vspace{1cm} (b)\hfill}
\end{figure}

\begin{figure}[p]
\includegraphics[height=4.92cm,width=6cm]{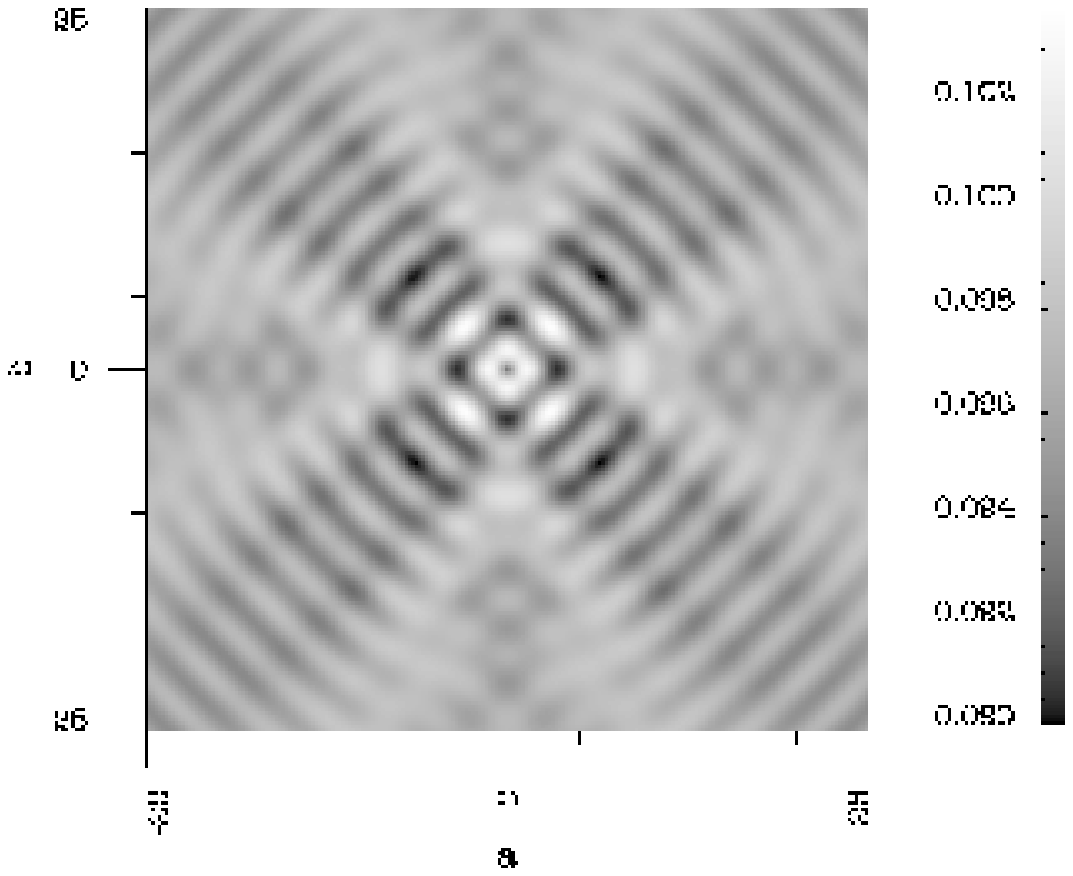}
\parbox{1cm}{\hfill\vspace{1cm} (c)\hfill}
\vspace*{0.5cm}
\caption{LDOS-map at the frequency $\omega=0.1\Delta$ around the Born impurity
located at (0,0) in the ab plane of the d-wave superconductor for
a) isotropic impurity potential;
b) potential in phase with the order parameter, $f({\bf k})=sgn(k^2_x-k^2_y)$;
c) potential out of the order parameter phase, $f({\bf k})=sgn(k_xk_y)$.
Units as in Fig. 4.}
\end{figure}


\begin{thebibliography}{20}
\bibitem{1} L. S. Borkowski and P. J. Hirschfeld, Phys. Rev. B {\bf 49},
15404 (1994).
\bibitem{2} A. M. Martin, G. Litak, B. L. Gy\"{o}rffy, J. F. Annett, and
K. I. Wysoki\'nski, Phys. Rev. B {\bf 60}. 7523 (1999).
\bibitem{3} C. O'Donovan and J. P. Carbotte, Phys. Rev. B {\bf 52}, 4568
(1995).
\bibitem{4} S. H. Pan, E. W. Hudson, K. M. Lang, H. Eisaki, S. Uchida,
and J. C. Davis, Nature {\bf 403}, 746 (2000).
\bibitem{5} E. W. Hudson, K. M. Lang, V. Madhavan, S. H. Pan, H. Eisaki,
S. Uchida, and J. C. Davis, Nature {\bf 411}, 920 (2001).
\bibitem{6} J. M. Byers, M. E. Flatt\'e, and D. J. Scalapino, Phys. Rev. Lett.
{\bf 71}, 3363 (1993).
\bibitem{7} M. I. Salkola, A. V. Balatsky, and D. J. Scalapino, Phys. Rev. Lett.
{\bf 77}, 1841 (1996).
\bibitem{8} J. Giapintzakis, D. M. Ginsberg, M. A. Kirk and S. Ockers,
Phys. Rev. {\bf B 50}, 15967 (1994).
\bibitem{9} S. Tolpygo, J. -Y. Lin, M. Gurvitch, S. Y. Hou, and J. M. Phillips,
Phys. Rev. {\bf B53}, 12454 (1996).
\bibitem{10} S. Tolpygo, J. -Y. Lin, M. Gurvitch, S. Y. Hou, and J. M. Phillips,
Phys. Rev. {\bf B53}, 12462 (1996).
\bibitem{11} J. -Y. Lin, S. J. Chen, S. Y. Chen, C. F. Chang, H. D. Yang,
S. K. Tolpygo, M. Gurvitch, Y. Y. Hsu, and H. C. Ku, Phys. Rev. B {\bf 59},
6047 (1999).
\bibitem{12} G. Hara\'n and A. D. S. Nagi, Phys. Rev. B {\bf 54}, 15463 (1996).
\bibitem{13} G. Hara\'n and A. D. S. Nagi, Phys. Rev. B {\bf 58}, 12441 (1998).
\bibitem{14} G. Hara\'n and A. D. S. Nagi, Phys. Rev. B {\bf 63}, 012503 (2001).
\bibitem{15} G. Hara\'n and A. D. S. Nagi, Acta Phys. Pol. B {\bf 32}, 3459
(2001).
\bibitem{15a} G. Hara\'n, Phys. Rev. B {\bf 65}, 216501 (2002).
\bibitem{16} C. H. Choi, Phys. Rev. B {\bf 63}, 064507 (2001).
\bibitem{17} P. Schlottmann, Phys. Rev. B {\bf 13}, 1 (1976).
\bibitem{18} A. P. Kampf and T. P. Devereaux, Phys. Rev. B {\bf 56}, 2360 (1997).
\end{thebibliography}
\end{document}